\documentclass[10pt,aps,reprint,prb,floatfix,superscriptaddress,longbibliography]{revtex4-2}
\pdfoutput=1
\usepackage{graphicx,latexsym}
\usepackage{dcolumn,amsfonts}
\usepackage{amssymb,amsmath,bm}

\usepackage[breaklinks=true,pdfencoding=auto]{hyperref}

\usepackage{natbib}
\usepackage{balance}

\hypersetup{
	colorlinks   = true, %Colours links instead of ugly boxes
	urlcolor     = blue, %Colour for external hyperlinks
	linkcolor    = blue, %Colour of internal links
	citecolor    = red %Colour of citations
}

\usepackage{color}
\usepackage{ulem}
\begin{document}

%-----------------------------------------------------------------

\title{Signatures of broken symmetries in the excitations of a
		periodic 2DEG\\ coupled to a cylindrical photon cavity}

\author{Vidar Gudmundsson}
\email{vidar@hi.is}
\affiliation{Science Institute, University of Iceland, Dunhaga 3, IS-107 Reykjavik, Iceland}
\affiliation{Department of Engineering, Reykjavik University, Menntavegur 1, IS-102 Reykjavik, Iceland}
\author{Vram Mughnetsyan}
\email{vram@ysu.am}
\affiliation{Department of Condensed Matter Physics, Yerevan State University, Alex Manoogian 1, 0025 Yerevan, Armenia}
\author{Hsi-Sheng Goan}
\email{goan@phys.ntu.edu.tw}
\affiliation{Department of Physics and Center for Theoretical Physics, National Taiwan University, Taipei 106319, Taiwan}
\affiliation{Center for Quantum Science and Engineering, National Taiwan University, Taipei 106319, Taiwan}
\affiliation{Physics Division, National Center for Theoretical Sciences, Taipei 106319, Taiwan}
\author{Jeng-Da Chai}
\email{jdchai@phys.ntu.edu.tw}
\affiliation{Department of Physics and Center for Theoretical Physics, National Taiwan University, Taipei 106319, Taiwan}
\affiliation{Center for Quantum Science and Engineering, National Taiwan University, Taipei 106319, Taiwan}
\affiliation{Physics Division, National Center for Theoretical Sciences, Taipei 106319, Taiwan}
\author{Nzar Rauf Abdullah}
\email{nzar.r.abdullah@gmail.com}
\affiliation{Physics Department, College of Science, University of Sulaimani, Kurdistan Region, Iraq}
\author{Chi-Shung Tang}
\email{cstang@nuu.edu.tw}
\affiliation{Department of Mechanical Engineering, National United University, Miaoli 360302, Taiwan}
\author{Wen-Hsuan Kuan}
\email{wenhsuan.kuan@gmail.com}
\affiliation{Department of Applied Physics and Chemistry, University of Taipei, Taipei 10048, Taiwan}
\author{Valeriu Moldoveanu}
\email{valim@infim.ro}
\affiliation{National Institute of Materials Physics, PO Box MG-7, Bucharest-Magurele, Romania}
\author{Andrei Manolescu}
\email{manoles@ru.is}
\affiliation{Department of Engineering, Reykjavik University, Menntavegur 1, IS-102 Reykjavik, Iceland}

%
%----------------------------------------------------------------

\begin{abstract}
{In a two-dimensional electron gas (2DEG) in a periodic lateral superlattice subjected to an external
homogeneous magnetic field and in a cylindrical far-infrared photon cavity we search for effects
of broken symmetries: Static ones, stemming from the unit cell of the system, and the external
magnetic field together with the dynamic ones caused by the vector potential of the cavity promoting
magnetic types of transitions, and the chirality of the excitation pulse.
The Coulomb interaction of the electrons is described within density functional theory, but
the electron-photon interactions are handled by a configuration interaction formalism within
each step of the density functional approach, both for the static and the dynamic system.
In the dynamical calculations we observe weak chiral effects that change character as
the strength of the electron-photon interaction and the external magnetic field are increased.
From the analysis of the chiral effects we identify an important connection of the para-
and diamagnetic electron-photon interactions that promotes the diamagnetic interaction
in the present system when the interaction strength is increased.
Furthermore, the asymmetric potential in the unit cell of the square array activates collective
oscillation modes that are not present in the system when the unit cell has a higher symmetry.}
\end{abstract}

\maketitle
%
%----------------------------------------------------------------------------------------
%

\section{Introduction}
Regular arrays or lateral superlattices in GaAs heterostructures have been used
for several decades to explore and predict subtle
transport \cite{Alves89:8257,Fang90:10171,Vasiliadou95:R8658,Gerhardts91:5192,Gerhardts89:1173},
energy band structural \cite{Petschel93:239,Lu2016,Rotter96:4452,Silberbauer92:7355},
and optical phenomena in a two-dimensional electron gas (2DEG) of extraordinary
purity and mobility \cite{HEITMANN200237,Demel90:788a,Heitmann93:56}.
Recently, the 2DEG system has been modeled and measured
in far-infrared (FIR) microcavity
systems \cite{Zhang1005:2016,PhysRevB.105.205424}.

Symmetries are used in condensed matter and atomic physics to restrict size requirements for
the ``computational space'' and resources necessary to solve a particular problem on one
hand, and on the other hand, to facilitate the construction of a model of a physical system
after the analysis and organization of experimental results gathered from it.

An external magnetic field can be used to restrict the dynamical motion of an
electron system by breaking its symmetry in space and time, but at the same time
the strength of the magnetic field introduces a new tunable external parameter
that can be of value. Moreover the magnetic field tends to relax restriction
on which optical transitions are forbidden or allowed in a particular
system \cite{PhysRevLett.78.4705,PhysRevE.96.012160,PhysRevB.67.041305,PhysRevB.67.045311}.

Similar statements can be made about the use of an external electric field,
photon fields of different optical cavities \cite{Wang2022,Wang2024-cm}, material composition, or
the design and shape of lateral superlattices in semiconductor heterostructures.

In our previous calculations of the excitation properties of a 2DEG in
periodic lateral superlattice potentials and subjected to the TE$_{011}$ mode
of a circular cylindrical FIR-photon cavity
we have only considered unit cells describing quantum
dots \cite{PhysRevB.110.205301},
or rings \cite{PhysRevB.111.115304}, so the
unit cell has always had a potential with circular symmetry with respect to its center.
In these systems an initial amplitude modulation of the circular electric field close to the
center of the cavity, where the long wavelength limits holds, always leads to excitations
that do not exhibit center-of-mass or dipole oscillations.
Instead, oscillations of the orbital magnetization and the photon number have been predicted in the
calculations.

In the QED-DFT-TP formalism the electron Coulomb interaction is treated within
a quantum electrodynamics density functional theory (QED-DFT) formalism,
but the electron-photon interaction
is described by exact diagonalization (configuration interaction) within each
iteration step of the DFT formalism. This is accomplished by using a linear
functional basis that is a tensor product (TP) of the electron and the photon
states \cite{10.1063/5.0123909,PhysRevB.110.205301-2}.

Here, we explore what kind of excitations appear when the unit cell does not
have this simple circular symmetry. We select a potential without a reflection symmetry
along one of the axes in its plane. The external homogeneous
magnetic field breaks the time and the chiral symmetry of the system. Besides effects of
the asymmetric unit cell we explore subtle effects caused by the rotational
sense of the initial excitation applied to the system, or the chirality of the
excitation. We emphasize that the choice of embedding the 2DEG into a cylindrical
FIR cavity with a TE$_{011}$ mode activates magnetic transitions. This would not be
the case if the cavity field would be represented by a constant vector potential.
This together with our search for chiral effects highlights an important connection
between the para- and the diamagnetic electron-photon interactions that explain
the growing importance of the diamagnetic interaction as the effective electron-photon
interaction strength is increased.

The paper is organized as follows: In Sec.\ \ref{Model} the model is briefly described,
the results and discussion thereof are found in Sec.\ \ref{Results}, with the conclusions
drawn in Sec.\ \ref{Conclusions}. Appendix \ref{Tech-details} contains technical details
of the methodology used for the modeling.

\section{Model}
\label{Model}
\begin{figure*}[htb]
	\includegraphics[width=0.48\textwidth]{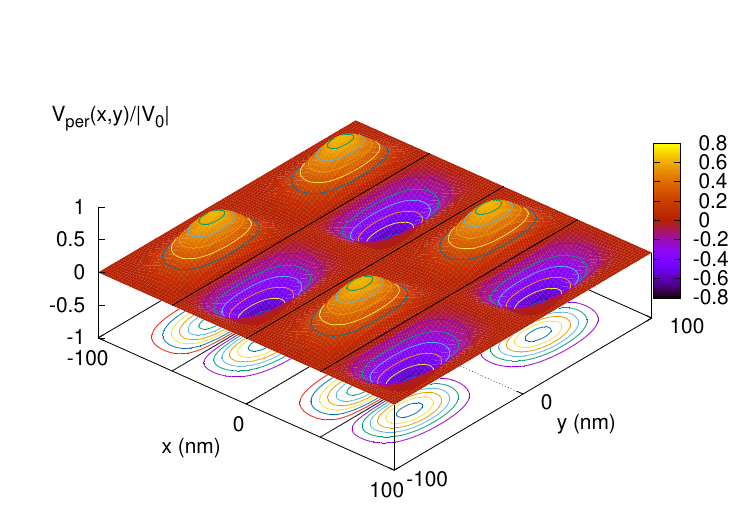}
	\includegraphics[width=0.48\textwidth]{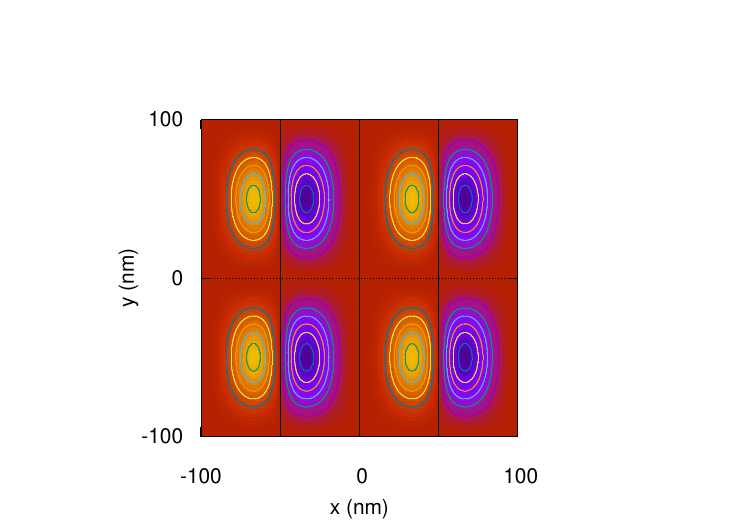}
	\caption{The periodic potential $V_\mathrm{per}(\bm{r})$ shown for 4
		neighboring unit cells in two different projections in configuration space.
		The same color scale is used in both panels
		to indicate the value of the potential.}
\label{FigVper}
\end{figure*}

The Hamiltonian of the 2DEG-cavity system is
\begin{equation}
	H = H_\mathrm{e} + H_\mathrm{int} + H_\gamma + H_\mathrm{ext}(t)
	\label{Htot}
\end{equation}
with the purely electron part
\begin{equation}
	H_\mathrm{e} = H_0 + H_\mathrm{Zee} + V_\mathrm{H} + V_\mathrm{per} + V_\mathrm{xc}
	\label{He}
\end{equation}
and the purely photon part
\begin{equation}
	H_\gamma = E_\gamma a^\dagger a \quad\mbox{with}\quad E_\gamma = \hbar\omega_\gamma
	\label{HEM}
\end{equation}
in terms of the photon creation, $a^\dagger$, and annihilation operator, $a$.
The asymmetric periodic lateral superlattice potential for the 2DEG in the
$xy$-plane
\begin{equation}
	V_\mathrm{per}(\bm{r}) = -V_0 \left\{\sin\left(\frac{g_1x}{2}\right)\sin\left(\frac{g_2y}{2}\right)\right\}^2
	                                   \sin\left(g_1x\right)
\label{Vper}
\end{equation}
is presented in Fig.\ \ref{FigVper}. In order to make the lack of
the left-right mirror symmetry in each unit cell clear, and the fact that
this choice of a unit cell is natural for $V_\mathrm{per}$ we show it
in two projections for 4 neighboring unit cells.
$V_0 = 8.0$ meV and the superlattice is spanned by the spatial vectors
$\bm{R}=n\bm{l}_1+m\bm{l}_2$ with $n,m\in \bm{Z}$,
and the unit vectors $\bm{l}_1 = L\bm{e}_x$ and $\bm{l}_2 = L\bm{e}_y$.
The reciprocal lattice is spanned by $\bm{G} = G_1\bm{g}_1 + G_2\bm{g}_2$ with
$G_1, G_2\in \mathbf{Z}$ and the unit vectors $\bm{g}_1 = 2\pi\bm{e}_x/L$
and $\bm{g}_2 = 2\pi\bm{e}_y/L$. The lattice length of the superlattice is $L = 100$ nm.

The direct Coulomb interaction expressed in terms of the local charge density
$\Delta n(\bm{r}) = n_\mathrm{e}(\bm{r})-n_\mathrm{b}$ is
\begin{equation}
	V_\mathrm{H}(\bm{r}) = \frac{e^2}{\kappa}\int_{\mathbf{R}^2}d\bm{r}'\frac{\Delta n(\bm{r}')}
	{|\bm{r}-\bm{r}'|}
	\label{Vcoul}
\end{equation}
with $-en_\mathrm{e}(\bm{r})$ the electron charge density and
$+en_\mathrm{b}$ the homogeneous positive background charge density
representing the ionized donors guaranteeing the charge neutrality of the total system.
The exchange correlation terms $V_\mathrm{xc}$ are detailed in Appendix A of
Ref.\ \cite{PhysRevB.110.205301} with references to their origins.
The total energy per unit cell of the 2DEG-cavity system is likewise given by
Eq.\ (25) in the same reference.
The external static vector potential ${\bm A} = (B/2)(-y,x)$ brings about the homogeneous
magnetic field $\bm{B}=B\hat{\bm{z}}$ perpendicular to the $xy$-plane containing the 2DEG.
The spin Zeeman Hamiltonian is
\begin{equation}
	H_\mathrm{Zee} = \pm g^* \mu_\textrm{B}^* B/2,
	\label{HZee}
\end{equation}
where $\mu_\textrm{B}^*$ is the effective Bohr magneton,
and the free electron Hamiltonian in terms of the generalized
momentum $\bm{\pi}$ is
\begin{equation}
	H_0 = \frac{1}{2m^*}\bm{\pi}^2, \quad\mbox{with}\quad
	\bm{\pi} = \left(\bm{p}+\frac{e}{c}\bm{A} \right).
	\label{H0}
\end{equation}
The external magnetic field yields the natural length scale, the magnetic length
$l = (\hbar c/(eB))^{1/2}$, and the energy scale, the cyclotron energy $\hbar\omega_c = (eB/(m^*c))$.

The electrons of the 2DEG interact with the photons of the cavity mode
\begin{align}
    H_\mathrm{int} = \frac{1}{c}\int_{\mathbf{R}^2} d\bm{r}\; &
    {\bm J}({\bm r})\cdot{\bm A}_\gamma (\bm{r}) \nonumber\\
    +& \frac{e^2}{2m^*c}\int_{\mathbf{R}^2} d\bm{r}\;
    n_\mathrm{e}(\bm{r})A^2_\gamma(\bm{r})
    \label{e-g}
\end{align}
with both the para- and the diamagnetic parts of the electron-photon coupling
included.

We follow a quantum electrodynamical density functional theory approach, QED-DFT-TP as presented by Malave
\cite{10.1063/5.0123909} and adapted to our 2DEG-cavity system by calculating
the energy spectrum and the eigenstates of the static part of the total Hamiltonian
(\ref{Htot}) in a linear functional basis built as a tensor product (TP) of the electron and photon states
\begin{equation}
	|\bm{\alpha\theta}\sigma n\rangle = |\bm{\alpha\theta}\sigma\rangle\otimes|n\rangle.
	\label{TP}
\end{equation}
The photon states $|n\rangle$ are the eigenstates of the photon number operator, and the electron states $|\bm{\alpha\theta}\sigma \rangle$ are the single-electron eigenstates of $H_0$
proposed by Ferrari and constructed
for the periodic 2DEG in an external magnetic field at each point in the first Brillouin zone, i.e.\ $\bm{\theta} = (\theta_1,\theta_2)\in [-\pi,\pi]\times[-\pi,\pi])$
\cite{Ferrari90:4598,Silberbauer92:7355,Gudmundsson95:16744,PhysRevB.105.155302,PhysRevB.106.115308}. $\sigma\in\{\uparrow,\downarrow\}$ is the quantum number for the $z$-component of the electron spin.
All further quantum numbers of the Ferrari states \cite{Ferrari90:4598} are combined in $\bm{\alpha}$, which can be construed as a subband index of the split Landau
bands \cite{Gudmundsson95:16744,Hofstadter76:2239,Dahl90:5763,Silberbauer92:7355,Pfannkuche92:12606}.

We note the Ferrari states with added spin information as $|\bm{\alpha\theta}\sigma\rangle$ and
the corresponding wavefunctions as $\phi_{\bm{\alpha\theta}\sigma}(\bm{r})$, but the states of the static total
Hamiltonian obtained after successfully converged iterations of the DFT Coulomb interacting electrons with
the electron-photon interactions included in each iteration step are noted by
$|\bm{\alpha\theta}\sigma n)$ with corresponding wavefunctions $\psi_{\bm{\alpha\theta}\sigma n}(\bm{r})$.

Most of the needed matrix elements are analytically known in the Ferrari basis so conveniently
the Liouville-von Neumann equation for the density operator
$\rho^{\bm{\theta}}$ at each $\bm{\theta}$ in reciprocal space
\begin{equation}
	i\hbar\partial_t \rho^{\bm{\theta}} (t) = \left[ H[\rho^{\bm{\theta}}(t)], \rho^{\bm{\theta}}(t)\right]
	\label{L-vN}
\end{equation}
will be solved there. In the $\{|\bm{\alpha\theta}\sigma)\}$-basis the initial conditions for the
corresponding density operator are given by the population (occupation) of the static equilibrium states
\begin{equation}
	r^{\bm{\theta}}_{\bm{\alpha}\sigma,{\bm{\beta}}\sigma'}(0) = f\left(E_{\bm{\alpha\theta}\sigma} - \mu \right)
	\delta_{\bm{\alpha},\bm{\beta}}\delta_{\sigma,\sigma'},
	\label{rho_0_int}
\end{equation}
where $f$ denotes the equilibrium Fermi distribution,
and the initial conditions in the $\{|\bm{\alpha\theta}\sigma\rangle\}$ basis become
\begin{equation}
	\rho^{\bm{\theta}}(0) = W^{\bm{\theta}}\left\{ r^{\bm{\theta}}(0)\right\} (W^{\bm{\theta}})^\dagger
	\label{rho_0}
\end{equation}
with $W^{\bm{\theta}}$ being the unitary transformation between
the two bases in each point $\bm{\theta}$ of the first Brillouin zone.
The expressions for the dynamical electron and current
densities are thus
\begin{align}
    n_\sigma (\bm{r},t)& =\nonumber\\
    \frac{1}{(2\pi)^2}&\int_{-\pi}^\pi d\bm{\theta}\,
    \sum_{\bm{\alpha\beta}}\phi^*_{\bm{\alpha\theta}\sigma}(\bm{r})
    \phi_{\bm{\beta\theta}\sigma}(\bm{r})\rho^{{\bm{\theta}}}_{\bm{\alpha}\sigma,\bm{\beta}\sigma}(t)
    \label{Net}
\end{align}
\begin{align}
    \bm{J}_{i}(\bm{r},t) = \frac{-e}{m^*(2\pi)^2}\sum_{\bm{\alpha\beta}\sigma}\int_{-\pi}^{\pi} d\bm{\theta}\;
    \Re&\left\{ \phi_{\bm{\alpha\theta}\sigma}^*(\bm{r})\bm{\pi}_i \phi_{\bm{\beta\theta}\sigma}(\bm{r}) \right\}\nonumber\\
    &\rho^{\bm{\theta}}_{\bm{\beta\theta}\sigma,\bm{\alpha\theta}\sigma}(t)
    \label{currD}
\end{align}            %%%%% Hér vantar inngang um grunna og ummyndun milli þeirra!!!!!!!!
respectively, for $i=x$ or $y$, and the total electron density is $n_\mathrm{e} = n_\uparrow + n_\downarrow$
in terms of the electron spin densities.

In the Appendix of Ref.\ \cite{PhysRevB.109.235306}
the vector potential of the cylindrical TE$_{011}$ cavity photon mode in the long wavelength
approximation is derived as
\begin{equation}
	{\bm A}_\gamma (\bm{r}) = \bm{e}_\phi {\cal A}_\gamma\left(a^\dagger_\gamma + a_\gamma  \right)
	\left(\frac{r}{l} \right),
\label{Ag}
\end{equation}
where $\bm{e}_\phi$ is the unit angular vector in the polar coordinates.
Importantly, this vector potential has the same spatial form as the static vector potential
${\bm A}$ leading to the
external homogeneous magnetic field $\bm{ B} = B\bm{e}_z$ \cite{PhysRevB.109.235306}.
Moreover, as (\ref{Ag}) depends on the spatial coordinate $r$ it promotes
magnetic transitions for the electrons. Below we will discuss how in the present system
it guarantees a connection between the para- and the diamagnetic electron-photon interaction
that increases the importance of the diamagnetic interaction as the electron-photon coupling and the external
magnetic field are increased. The spatially dependent vector potential (\ref{Ag})
of the electron-photon interaction (\ref{e-g}) means that the paramagnetic part of the
interaction can not be transformed to an electron dipole interaction, and the diamagnetic part
can not be approximated by a constant depending on the number of electron.

In terms of the photon creation and annihilation operators,
$a^\dagger_\gamma$ and $a_\gamma$, the electron-photon interaction (\ref{e-g}) becomes
\begin{align}
	H_\mathrm{int} &= g_\gamma \hbar\omega_c \left\{ lI_x + lI_y\right\} \left(a^\dagger_\gamma + a_\gamma\right)\nonumber\\
	&+ g^2_\gamma \hbar\omega_c {\cal N}\left\{\left(a^\dagger_\gamma a_\gamma + \frac{1}{2}\right)
	+\frac{1}{2}\left(a^\dagger_\gamma a^\dagger_\gamma + a_\gamma a_\gamma\right)\right\}
	\label{e-gIxIyN}
\end{align}
with the integrals, $I_x$, $I_y$, and ${\cal N}$, that can be interpreted as functionals
of the electron current and charge densities, respectively, defined as \cite{PhysRevB.109.235306}
\begin{align}
	l(I_x + I_y) &=  \frac{m^*}{e}\int_{\bm{{\cal A}}} d\bm{x}\;
	\frac{l}{\hbar}\left[-J_x(\bm{x})\left(\frac{y}{l}\right) + J_y(\bm{x})\left(\frac{x}{l}\right)\right]
	\nonumber\\
	{\cal N} &= \int_{\bm{{\cal A}}} d\bm{x}\; n_\mathrm{e}(\bm{x})\left(\frac{x^2+y^2}{l^2}\right).
\label{lIxx}
\end{align}
All of the terms of the electron-photon interaction, also the ones often designed as the
antiresonant ones of the diamagnetic electron-photon interaction are included in the calculation
and in the present system they are of paramount
importance, just like was seen for the dynamic results for the array of
quantum dots \cite{PhysRevB.110.205301} and rings \cite{PhysRevB.111.115304}
in a cylindrical photon cavity with a TE$_{011}$ mode.
The dimensionless electron-photon coupling constant is
\begin{equation}
	g_\gamma = \left\{ \left( \frac{e{\cal A}_\gamma}{c} \right) \frac{l}{\hbar} \right\},
\end{equation}
and the energy of the photons is $E_\gamma = \hbar\omega_\gamma$.

The real-time excitation of the 2DEG-cavity system is set up by applying a
short time-dependent modulation of the electron-photon coupling described by the
time-dependent Hamiltonian
\begin{align}
	H_\mathrm{ext}(t) &= F(t)
	\biggl[ g_\gamma \hbar\omega_c\left\{lI_x + lI_y \right\} \left( a^\dagger + a \right)\nonumber\\
	+ &\left. g_\gamma^2 \hbar\omega_c {\cal N} \left\{ \left( a^\dagger a + \frac{1}{2}\right)
	+ \frac{1}{2}\left( a^\dagger a^\dagger + a a  \right) \right\} \right]
	\label{Ht}
\end{align}
with
\begin{equation}
	F(t) = \left( \frac{V_t}{\hbar\omega_c}\right) (\Gamma t)^2 \exp{(-\Gamma t)}\cos{(\omega_\mathrm{ext} t)}.
	\label{ft}
\end{equation}
$\omega_\mathrm{ext}$ is the frequency of the modulation of the electron-photon interaction.

We emphasize that the time-dependent Hamiltonian in the Liouville-von Neumann equation (\ref{L-vN})
is a functional of the density operator
$\rho^{\bm{\theta}}$ through the electron charge and current densities, that have to be
updated in each iteration within each time-step of the numerical time integration of (\ref{L-vN}).
The Crank-Nicolson scheme is used as it is designed for use in Hermitian systems \cite{Crank1947}.

Besides the dynamic electron density and the current density the calculated time-dependent
density matrix is used to evaluate the time-dependent average photon number
\begin{equation}
    N_\gamma (t) = \frac{1}{(2\pi)^2}\sum_{\sigma}\int^{\pi}_{-\pi} d\bm{\theta}\; \mathrm{Tr} \left\{ \rho^{\bm{\theta}}_{\sigma}(t) a^\dagger_\gamma a_\gamma\right\},
    \label{Ngt}
\end{equation}
and the dynamical orbital magnetization
to monitor current oscillations \cite{Desbois98:727,Gudmundsson00:4835}
\begin{equation}
    Q_J(t) = \frac{1}{2c{\cal A}}\int_{\cal A} d\bm{r} \left( {\bf r}\times
    \langle {\bf J}({\bf r},t) \rangle \right) \cdot{\bm{e}_z},
    \label{Mo}
\end{equation}
with ${\cal A} = L^2$. The number of electrons in each unit cell is noted by $N_\mathrm{e}$.
In addition to observe effects of the asymmetry of the
unit cell we calculate the mean values, $Q_0 = \langle x^2+y^2 \rangle$, $Q_2 = \langle xy\rangle$,
and $Q_1 = \langle x\rangle$ and $\langle y\rangle$. These mean values are calculated
directly from the electron time-dependent density and supply information about,
breathing or monopole density oscillations, quadrupole modes, and dipole modes excited in the
system, respectively.

%------------------------
\section{Results}
\label{Results}

In the calculations, we assume GaAs parameters: the effective electron mass
$m^* = 0.067m_e$, the dielectric constant $\kappa = 12.4$, and the effective
$g$-factor $g^* = -0.44$.
The magnetic flux through a unit cell is in accordance with the commensurability condition
$B{\cal A}= BL^2 =pq\Phi_0$ with $p$ and $q$ integers
\cite{Ferrari90:4598,Silberbauer92:7355,Gudmundsson95:16744},
and the magnetic flux quantum $\Phi_0 = hc/e$.
Below, we use $pq = 1$ corresponding to $\hbar\omega_c = 0.7145$ meV and $pq = 2$ corresponding to
$\hbar\omega_c = 1.429$ meV to indicate the cases of {\lq\lq}low{\rq\rq} and {\lq\lq}high{\rq\rq}
magnetic fields. In the static calculations we assume the temperature $T = 1$ K.

\begin{figure}[htb]
    \includegraphics[width=0.48\textwidth]{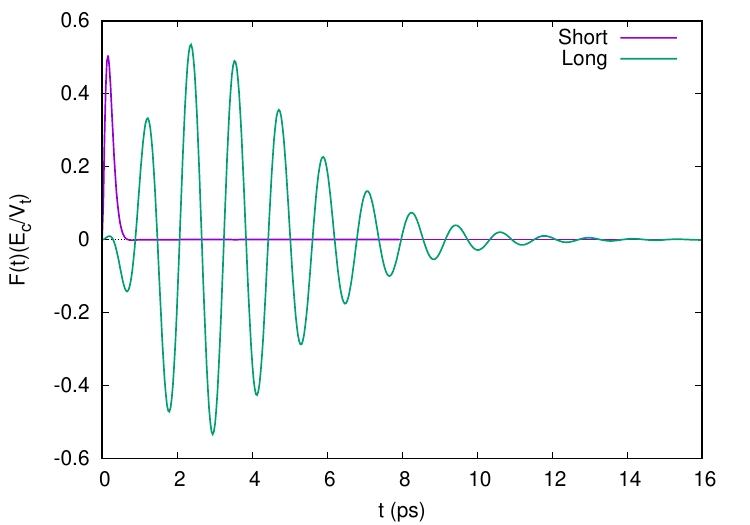}
    \caption{The temporal dependence of the excitation of the 2DEG-cavity system
             $F(t)$ for the cases of a short or long excitation pulse.
             For the long pulse $\hbar\omega_\mathrm{ext} = 3.5$ meV, and $\hbar\Gamma = 0.5$ meV,
             but for the short pulse $\hbar\omega_\mathrm{ext} = 1.5$ meV, $\hbar\Gamma = 8.0$ meV.
             Specially note that for the case of a short pulse the sign of $V_t$ defines the
             chirality of the pulse.  $E_c = \hbar\omega_c$.}
\label{pulses}
\end{figure}

We use two different excitation pulses (\ref{ft}), one labeled as {\lq\lq}long{\rq\rq} with
$\hbar\omega_\mathrm{ext} = 3.5$ meV, $\hbar\Gamma = 0.5$ meV, and the
other labeled as {\lq\lq}short{\rq\rq} with $\hbar\omega_\mathrm{ext} = 1.5$ meV, $\hbar\Gamma = 8.0$ meV.
They are compared in Fig.\ \ref{pulses}.
The shorter pulse is kept short enough for its impulse to be felt only in
one sense or direction of rotation, i.e.\ it has a clear chirality defined by the sign of $V_t$,
but the other longer one causes oscillations in both directions, though as will be seen later,
it delivers a slightly more impulse in one direction of rotation.
Instead of an impulse it is more appropriate here to state that the
excitation of the system supplies angular momentum to it, and due to the
effects of the static external magnetic field the system is differently
susceptible to this added or subtracted angular momentum.
The time step in the integration of the Liouville-von Neumann equation (\ref{L-vN})
is chosen to be $\Delta t = 0.02$ ps and the $(\Gamma t)^2$ factor in the excitation
modulation $F(t)$ is chosen in order to have the onset of the excitation smooth.

\begin{figure*}[htb]
    \includegraphics[width=0.34\textwidth]{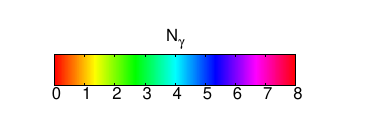}\\
    \vspace*{-0.3cm}
    \includegraphics[width=0.30\textwidth]{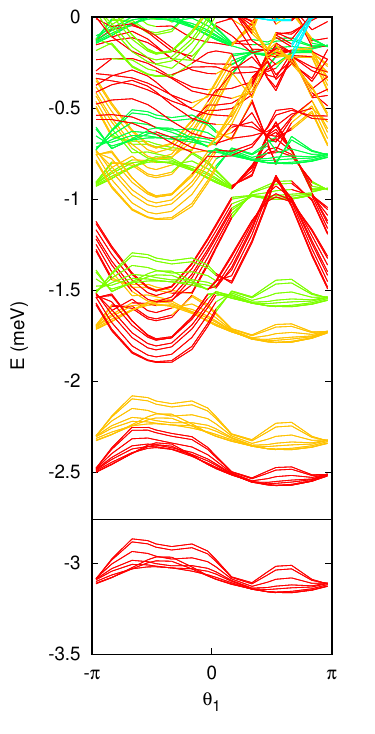}
    \includegraphics[width=0.30\textwidth]{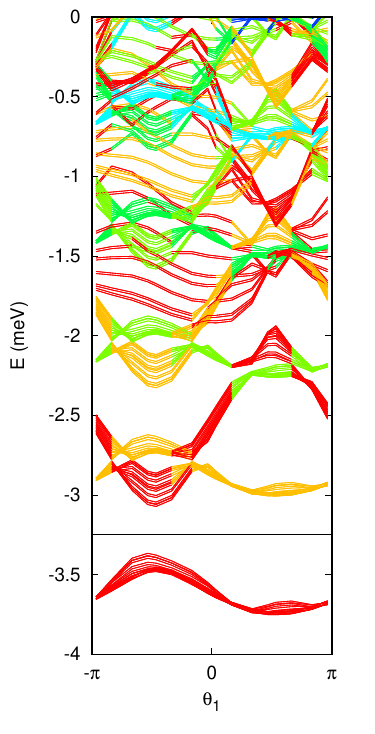}
    \includegraphics[width=0.30\textwidth]{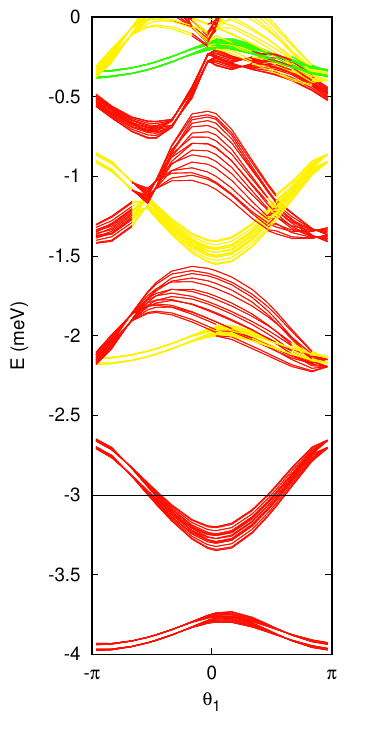}
    \caption{The energy bandstructure for $N_\mathrm{e}=1$, $pq=1$, $E_\gamma =0.7$ meV,
             and $g_\gamma =0.08$ (left), $N_\mathrm{e}=2$, $pq=1$, $E_\gamma =0.7$ meV,
             and $g_\gamma =0.04$ (center), and $N_\mathrm{e}=3$, $pq=2$, $E_\gamma =1.0$ meV,
             and $g_\gamma =0.08$ (right) projected on the $\theta_1$ direction in reciprocal
             space. The chemical potential is indicated by a horizontal black line.
             The photon content of the energy bands is color coded according to the
             color bar at top.}
\label{E-spectra}
\end{figure*}

Figure \ref{E-spectra} presents the energy band structure projected on the
$\theta_1$ direction in the first Brillouin zone for $N_\mathrm{e}=$ 1,2, and 3 electrons
in the unit cell for parameters that are relevant for the discussion of
further results below.
The chemical potential $\mu$ is indicated with a black horizontal line in each
subfigure. The photon content of each Landau subband is color coded in the figure, and
it is, for example, possible to locate the first photon replica of the lowest band higher up in
the spectra. Projection of the energy spectra on the other direction in reciprocal space, $\theta_2$,
is not shown, as it leads to totally symmetric bands around $\theta_2 = 0$. The asymmetry of the bands comes from
the asymmetry of the unit cell in one spatial direction.  For $N_\mathrm{e}=1$ (left) a large
spin splitting of the lowest bands is seen, the chemical potential is located between the
two lowest spin bands. For $N_\mathrm{e}=2$ only the tiny spin splitting expected for the small
g-factor of GaAs is present and the two lowest occupied spin bands overlap (center). Due to the
strong asymmetric spatial modulation of $V_\mathrm{per}$ the spin splitting is still
weak for $N_\mathrm{e}=3$ (right).

This is apparently in disagreement with older results on the g-factor enhancement in a
periodic 2DEG due to the Coulomb exchange effects \cite{PhysRevB.56.9707}. However,
in a previous calculation for an array of quantum rings
the Coulomb exchange effects at low electron-photon coupling were generally strong enough for $N_\mathrm{e}=3$
to cause a large spin splitting, that vanished for stronger electron-photon
coupling \cite{PhysRevB.111.115304}.

The mean photon content $N_\gamma$ of the unit cell of the system in Fig.\ \ref{E-spectra}
is 0.00797 (left), 0.0435 (center), and 0.264 (right) reflecting the fact that no rather pure photon
replica is occupied, but the electron photon coupling influences all subbands. The effective
{\lq\lq}coupling strength{\rq\rq} is thus low, though $g_\gamma$ may not be small.
The effective coupling strength depends also on the photon energy, the electron number,
the charge and current densities together with the band structure of the periodic
potential.
We note that specially for the lower magnetic field, $pq=1$, the higher order photon
replicas can be strongly transformed by the interaction of the electrons in neighboring
unit cells, i.e.\ for $pq=1$ the magnetic length is $l = 39.89$ nm and there is a considerable
overlapping of the electron density of neighboring unit cells.

\begin{figure*}[htb]
	\includegraphics[width=0.48\textwidth]{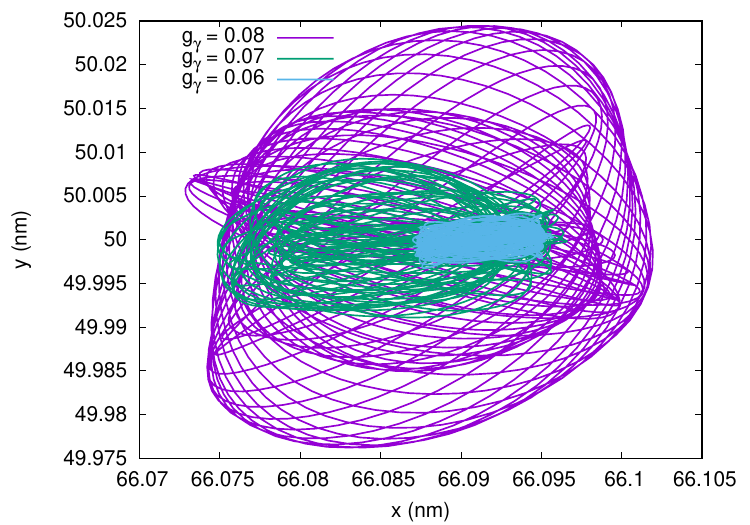}
	\includegraphics[width=0.48\textwidth]{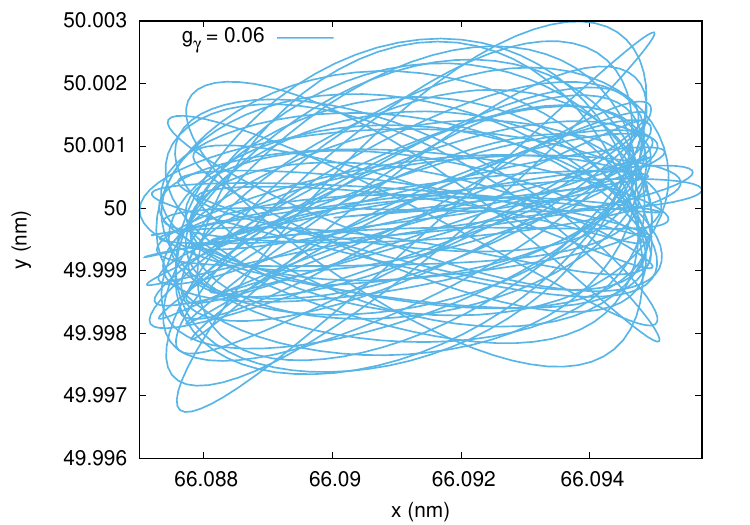}\\
	\includegraphics[width=0.48\textwidth]{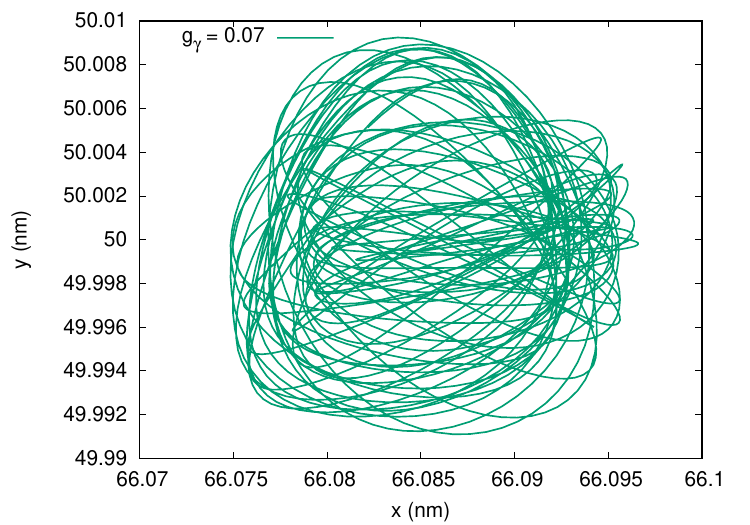}
	\includegraphics[width=0.48\textwidth]{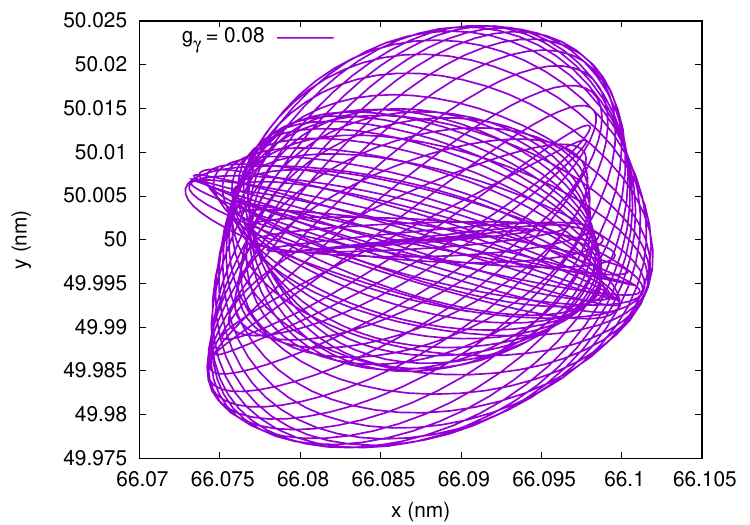}
	\caption{The mean values for the spatial coordinates of the center of mass
	         for the electron density within a unit cell for the first approximately
	         60 - 65 ps after the onset of the excitation pulse
	         for different values of the dimensionless electron-photon
	         coupling $g_\gamma$. $pq=2$, $N_\mathrm{e} = 3$, $V_t=0.1\hbar\omega_c$,
	         $\hbar\omega_c=1.429$ meV. The system is excited with the long
	         excitation pulse with $\hbar\omega_\mathrm{ext} = 3.5$ meV, and $\hbar\Gamma = 0.5$ meV.}
\label{CM-pq2}
\end{figure*}

Like was seen for the square array of quantum dots or rings the excitation
(\ref{Ht}) together with the TE$_{011}$ cavity mode (\ref{Ag}) does not promote
center of mass collective oscillations for the electron density, instead the
time dependent circular electric field of the cavity mode leads to changes in the
rotational currents that in turn cause changes in the mean photon number and the
orbital magnetization. The broken symmetry of the unit cell imposed by the
present periodic potential (\ref{Vper}) changes this situation and small dipole
or center of mass oscillations are observed as is seen in Fig.\ \ref{CM-pq2}.
Fig.\ \ref{CM-pq2} shows the paths of the paths the mean values $\langle x\rangle$ and  $\langle y\rangle$
of the electron density of one unit cell for approximately the first 60 ps from the start of the
excitation of the system at $t = 0$. We can interpret these as the paths of the center of mass and
in the upper left panel we see the paths for 3 different values of $g_\gamma$, the strength of
the electron-photon coupling for the case of $pq=2$, $N_\mathrm{e}=3$, and photon energy
$E_\gamma = 1.00$ meV. Each individual path for a certain coupling strength is then
repeated in the other 3 panels for clarity. Clearly, in addition to the different
spatial domain of the oscillations one notices immediately their increased regularity with
increasing electron-photon coupling. The change in the domains has to be explained in terms
of the enhanced polarizability of the electron charge density with increasing $g_\gamma$.
The regularity comes from the increasing dominance of the electron-photon coupling in comparison
with the Coulomb forces and the periodic potential. Below we will note that the dipole
oscillation mode is indeed much smaller than other modes excited in the system.

\begin{figure*}[htb]
	\includegraphics[width=0.48\textwidth]{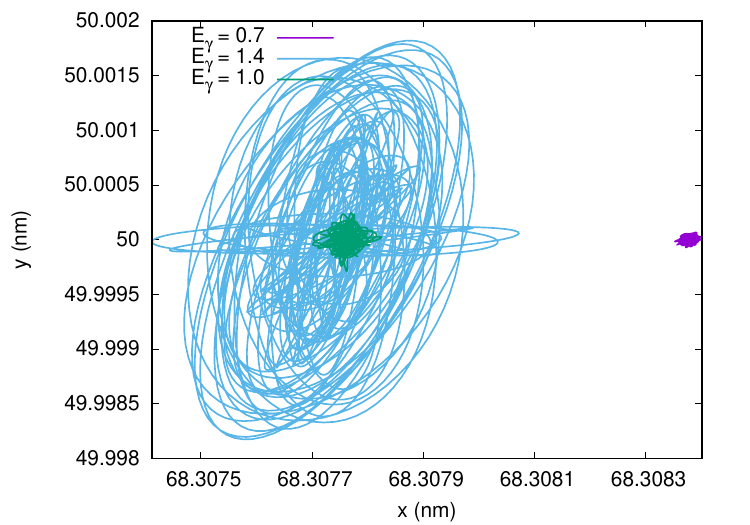}
	\includegraphics[width=0.48\textwidth]{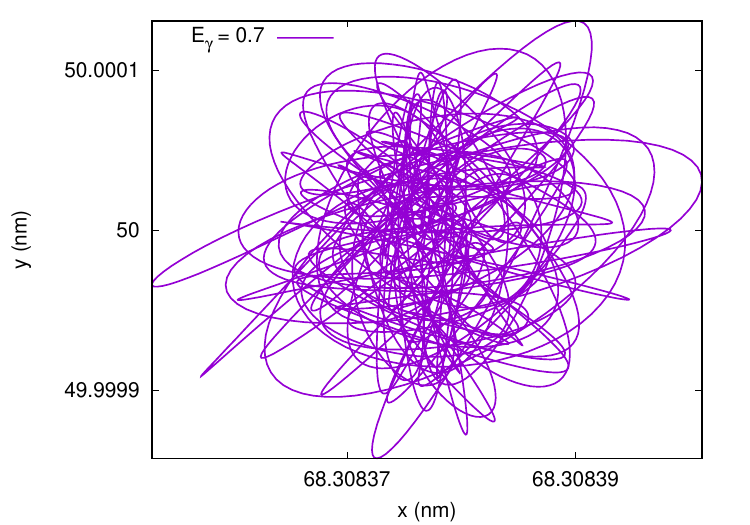}\\
	\includegraphics[width=0.48\textwidth]{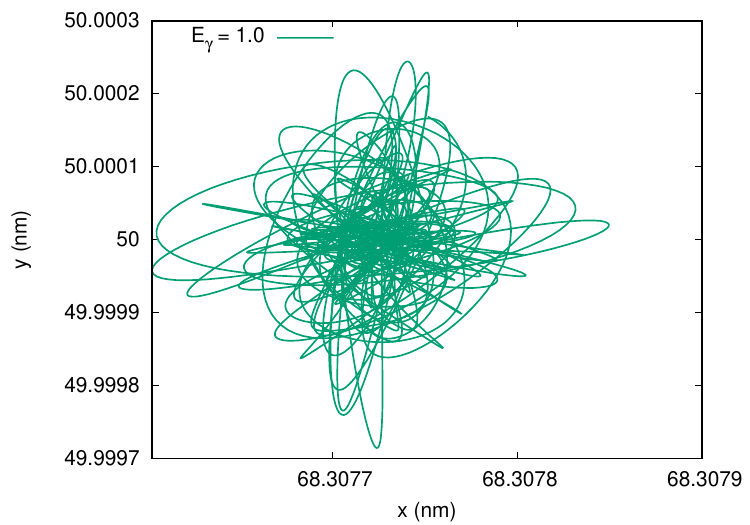}
	\includegraphics[width=0.48\textwidth]{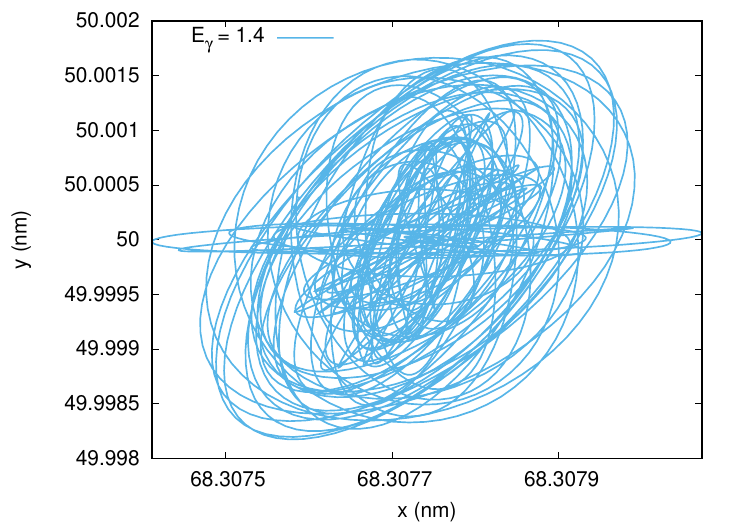}
	\caption{The mean values for the spatial coordinates of the center of mass
		for the electron density within a unit cell for the first
		100 ps after the onset of the excitation pulse
		for different values of the photon energy $E_\gamma$.
		$g_\gamma = 0.08$, $pq=1$, $N_\mathrm{e} = 1$, $V_t=0.1\hbar\omega_c$,
		$\hbar\omega_c=0.7145$ meV. The system is excited with the long
		excitation pulse with $\hbar\omega_\mathrm{ext} = 3.5$ meV, and $\hbar\Gamma = 0.5$ meV.}
\label{CM-pq1}
\end{figure*}

In Fig.\ \ref{CM-pq1} we display similar information about the center of mass oscillations
for the case of $pq=1$, $g_\gamma = 0.08$ and 3 different values of the photon energy
$E_\gamma$, but now for only one electron in each unit cell, $N_\mathrm{e}=1$.
Here, the paths are shown for 100 ps. We do not see the same regularity for any of the
cases here as was seen for the last panel in Fig.\ \ref{CM-pq2}. The magnetic field is
lower and even if there is just one electron in the unit cell it feels the presence of
electrons in the neighboring cells through the increased density overlap, or the
band structure at the lower magnetic field, see Fig.\ \ref{E-spectra}.
Moreover, as only the lowest Landau subband is occupied the effective coupling constant
is always low and thus the photon content remains very low.
The photon content is 0.00797 ($E_\gamma =0.7$), 0.00189 ($E_\gamma =1.0$), 0.000935 ($E_\gamma =1.4$),
so as expected the photon content (or the effective coupling) decreases with increasing photon energy.
The first photon replica is farther away from the lowest occupied Landau band as the photon
energy increases.
The increased oscillation amplitude observed in the last panel for $E_\gamma = 1.4$ meV
has to be understood in terms of the energy bandstructure of the system. The energy
bandstructure for $E_\gamma = 0.7$ meV is displayed in the left panel of Fig.\ \ref{E-spectra}.
There the first photon replica subband is located just above the mostly-unoccupied spin
subband above the chemical potential. When the photon energy is increased to $E_\gamma = 1.4$ meV
the first photon replica {\lq\lq}interacts{\rq\rq} with the broad 3rd electron subband,
similar as the second photon replica does in the left panel of Fig.\ \ref{E-spectra}.
The first photon replica is high enough in energy to interact with a Landau subband with large dispersion
indicating that it represents states that are not highly localized inside a unit cell
and the interaction creates regions of subband anticrossing marking underlying resonance conditions.
The interaction with the cavity photons with higher energy $E_\gamma$ increases the
delocalization of the electrons in the system. We have also calculated the motion of the
center of mass for the higher photon energy $E_\gamma = 1.6$ meV (not shown here), but otherwise with the same
parameters as in the last subfigure of Fig.\ \ref{CM-pq1}. The photon content at $E_\gamma = 1.6$
is slightly lower than at $E_\gamma = 1.4$ pointing to a lower effective coupling constant, but
the amplitude of the center of mass oscillations is a bit larger as the
first photon replica subband is still interacting with the same Landau subband, but at a
slightly higher energy.
Clearly, the amplitude of the center of mass oscillations, or the delocalization of the electrons,
is not a simple monotonous function of the photon energy and similarly the equilibrium center of mass
of the electrons is not a simple function of $E_\gamma$ and $g_\gamma$.

When comparing Fig.\ \ref{CM-pq2} and \ref{CM-pq1} we should have in mind that
in Fig.\ \ref{CM-pq2} there are 3 electrons in each unit cell, and Coulomb screening effects for the Li-like system can lead to high polarizability of the electron charge density in each cell.
In general, the amplitude of the center of mass oscillations is small here as the
strength of the initial excitation is kept small, $V_t/(\hbar\omega_c) = 0.1$.

\begin{figure*}[htb]
    \includegraphics[width=0.48\textwidth]{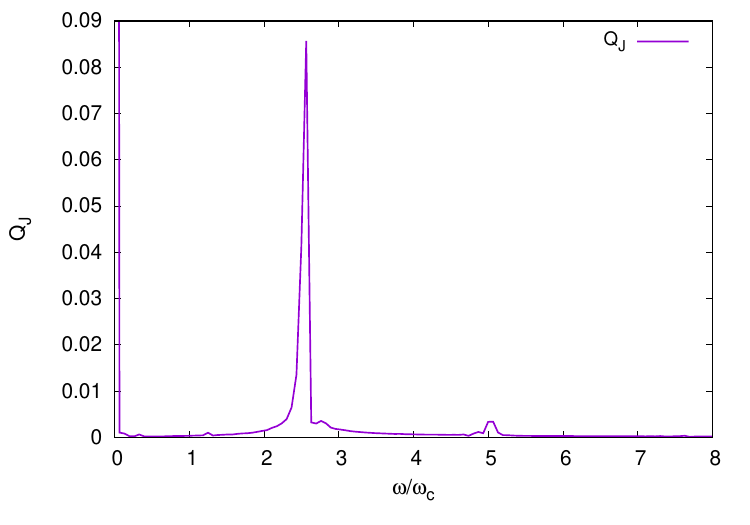}
    \includegraphics[width=0.48\textwidth]{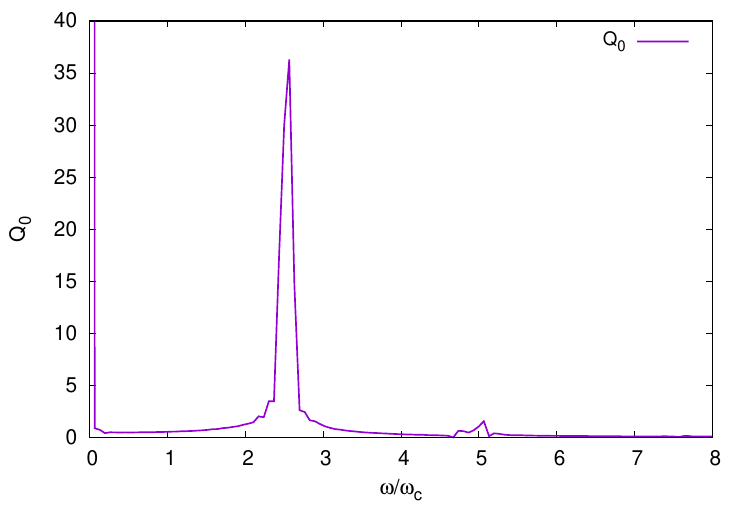}\\
    \includegraphics[width=0.48\textwidth]{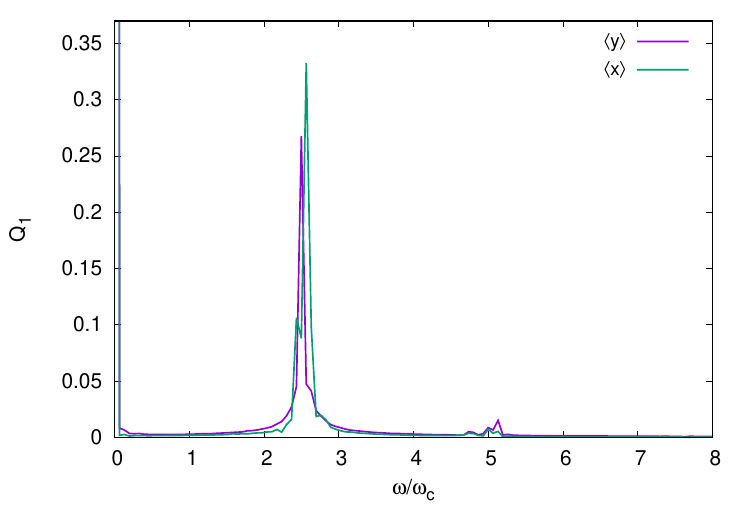}
    \includegraphics[width=0.48\textwidth]{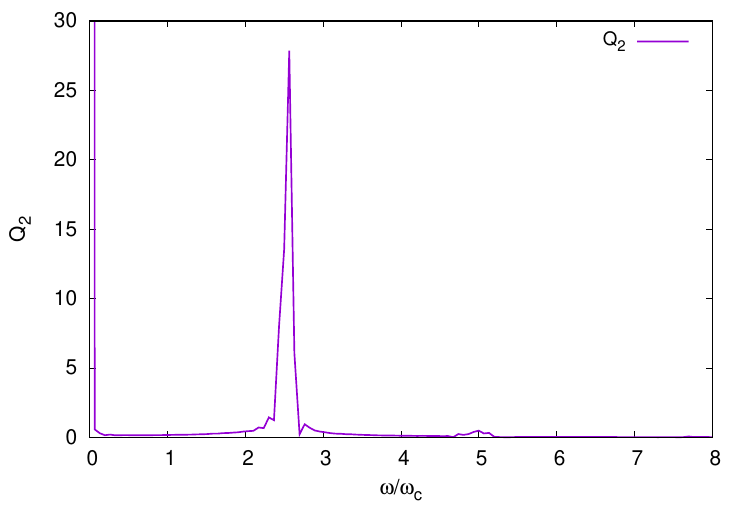}
    \caption{The Fourier power spectra for $Q_J$, $Q_0$, $Q_1$, $Q_2$.
             $pq=2$, $N_\mathrm{e}=3$, $g_\gamma=0.08$, $E_\gamma=1.00$ meV.
             The system is excited with the long
             excitation pulse with $\hbar\omega_\mathrm{ext} = 3.5$ meV, and $\hbar\Gamma = 0.5$ meV.}
\label{Qi-pq2}
\end{figure*}

Next we explore which collective oscillation modes are found in the system after excitation.
We start with the case of $N_\mathrm{e}=3$, $pq=2$, $g_\gamma=0.08$, and $E_\gamma=1.00$ meV
so, 3 electrons in a unit cell at the higher value of magnetic field $B=0.827$ T corresponding to
the cyclotron energy $\hbar\omega_c = 1.429$ meV. The system is excited with the
{\lq\lq}long{\rq\rq} excitation pulse shown in Fig.\ \ref{pulses}, and the Fourier power spectra
for the quantities $Q_J$, $Q_0$, $Q_1$, and $Q_2$ are displayed in Fig.\ \ref{Qi-pq2}.
The power spectra are calculated of the time series after the excitation pulse has vanished,
so from $t=16.0$ ps. The average photon number in the unit cell of the system for these parameters
is 0.264 at $t=0$ and the center of mass oscillations are very regular as was seen in the last panel
of Fig.\ \ref{CM-pq2}. The corresponding energy bandstructure is shown in the right panel of
Fig.\ \ref{E-spectra}. The average photon number and $g_\gamma = 0.08$ point to a rather strong
electron-photon interaction in the system. The power spectra for all the variables $Q_i$ show
a main peak at the same location as the only visible peak in the photon oscillation power
spectra is located at (not shown here). The main peak is at an energy of twice the cyclotron
energy plus a shift due to the strong electron-photon interaction around $\omega /\omega_c\sim 2.5$.
In addition a weak peak is seen at approximately double this energy. So, just like in the calculations
for the 2DEG in arrays of quantum dots or rings in a cylindrical FIR cavity with a
TE$_{011}$ photon mode the excited modes of the system are mostly caused by two photon modes of
the diamagnetic electron-photon interactions \cite{PhysRevB.110.205301,PhysRevB.111.115304}.
The low peaks representing the one-photon excitations
caused by the paramagnetic interaction are too low to be seen on the scales of Fig.\ \ref{Qi-pq2}, but
they can be found by using logarithmic vertical scales.

It is fair to point out that below, for a lower effective electron-photon
interaction, we will even have cases where the paramagnetic electron-photon interaction
dominates the power spectra of $Q_i$ as should be expected since the strength of that
interaction is proportional to $g_\gamma$, but the diamagnetic part of the interaction
is proportional to  $g^2_\gamma$. We also have to be clear about the excitation used
here, (\ref{Ht}) with (\ref{ft}). It can not directly lead to center of mass oscillation
as it does not couple different $\bm{\theta}$ points in the reciprocal space.
The small center of mass oscillations here are only a secondary effect caused by the
lack of reflection symmetry of $V_\mathrm{per}$. The peaks in $Q_1$ in Fig.\ \ref{Qi-pq2}
are thus not at energies that could be inferred from a linear response calculation
for the system subject to dipole excitation field, but interestingly, we see a
splitting of the main $Q_1$ peak for $\langle x\rangle$ and $\langle y\rangle$, that
can be understood in terms of the symmetry difference of $V_\mathrm{per}$ for these
two spatial directions.

\begin{figure*}[htb]
    \includegraphics[width=0.48\textwidth]{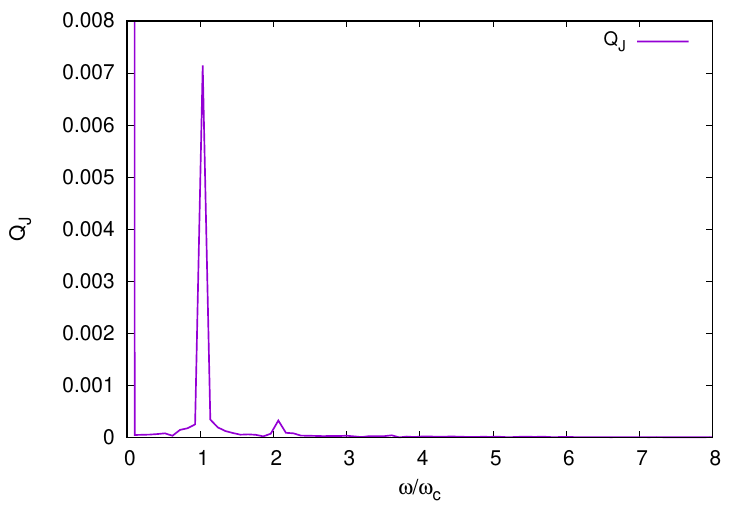}
    \includegraphics[width=0.48\textwidth]{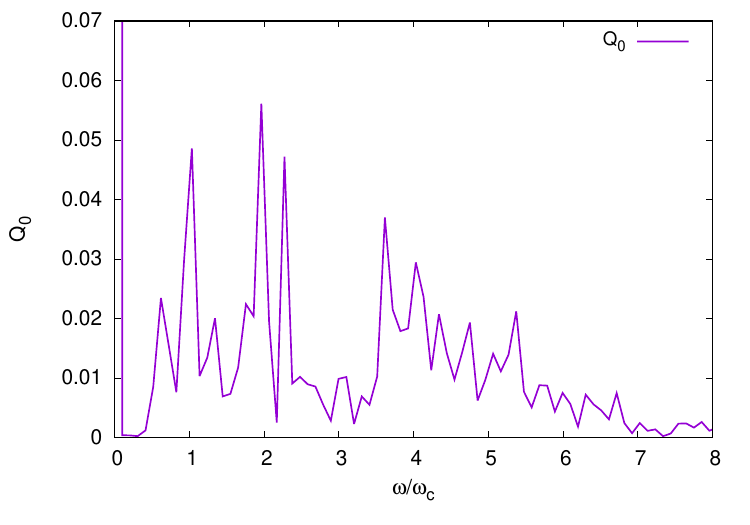}\\
    \includegraphics[width=0.48\textwidth]{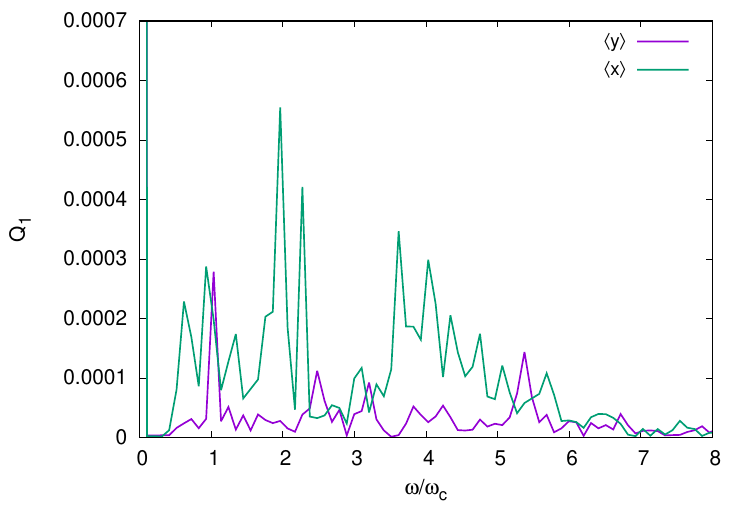}
    \includegraphics[width=0.48\textwidth]{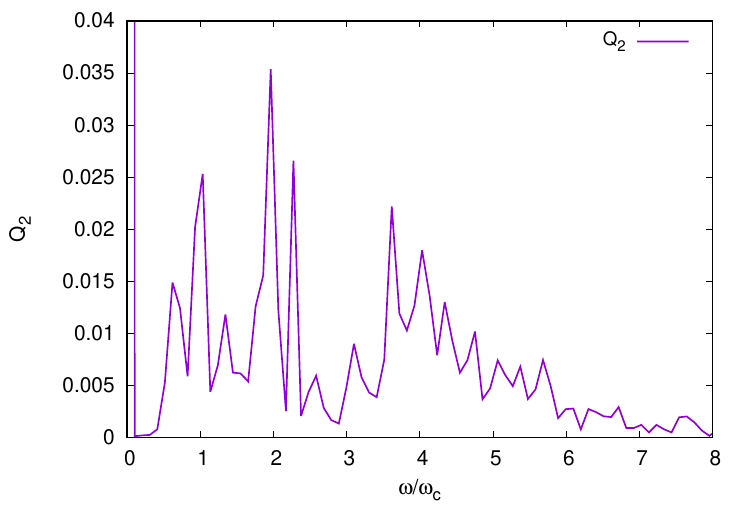}
    \caption{The Fourier power spectra for $Q_J$, $Q_0$, $Q_1$, $Q_2$.
             $pq=1$, $N_\mathrm{e}=2$, $g_\gamma=0.04$, $E_\gamma=0.70$ meV.
             The system is excited with the long
             excitation pulse with $\hbar\omega_\mathrm{ext} = 3.5$ meV, and $\hbar\Gamma = 0.5$ meV.}
\label{Qi-pq1}
\end{figure*}

For the system at the lower magnetic field $B=0.413$ T corresponding to $pq=1$ and for
$N_\mathrm{e}=2$ the calculated power spectra for $Q_i$ look very different as can be
seen in Fig.\ \ref{Qi-pq1}.
The reason is that for this low magnetic field the magnetic length $l=39.89$ nm and
even though the electron density has a maximum around the minimum of the potential
$V_\mathrm{per}$ it overlaps with the density from other unit cells and the energy
bandstructure bears clear signs of this. The Fourier power spectrum of $Q_J$ shows a
strong peak at approximately the cyclotron resonance energy caused by single photon
transitions of the paramagnetic electron-photon interaction with a lower peak
for the two-photon transitions of the diamagnetic interaction. The power spectrum for
the average photon number (not shown here) only shows these same two peaks, but the
strength of the second peaks is higher there even though the paramagnetic peaks is
still the main peak.

The power spectra for $Q_0$, $Q_1$, and $Q_2$ in Fig.\ \ref{Qi-pq1} are too complex to
allow for a precise analysis, but they reflect all kind of internal motion in the
electron density promoted by the particular shape and symmetry of $V_\mathrm{per}$.
The multitude of oscillations can be observed by the time dependent induced density.
Below, we will do that for a less complex case where the different modes can be separated.

\begin{figure*}[htb]
    \includegraphics[width=0.48\textwidth]{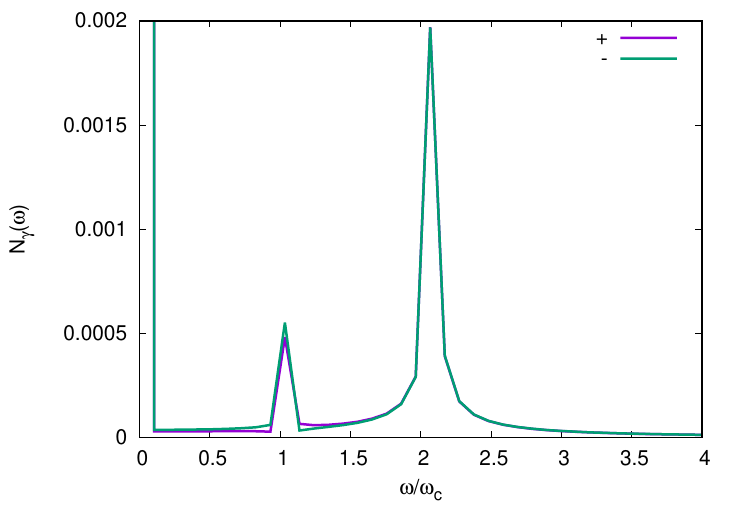}
    \includegraphics[width=0.48\textwidth]{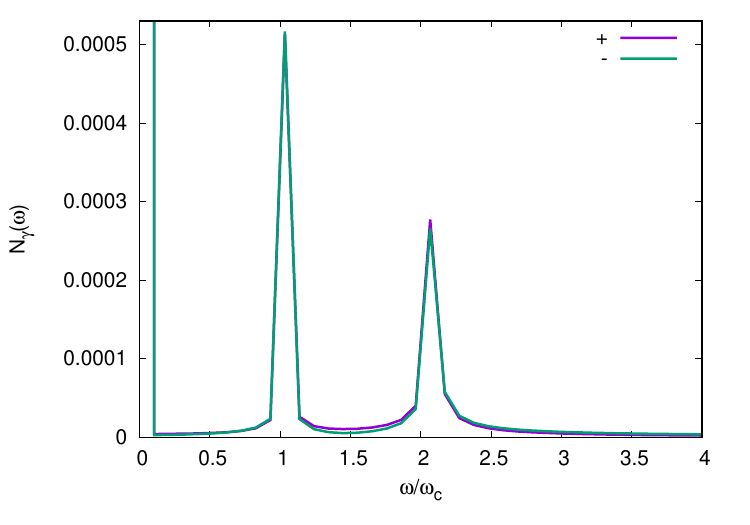}
    \caption{The Fourier power spectra for the mean photon number $N_\gamma (\omega )$
             for $V_t/\hbar\omega_c = \pm 0.1$
             for a short (left), and a long excitation (right). $pq=1$, $N_\mathrm{e}=2$, $g_\gamma=0.04$, $E_\gamma=0.70$ meV.}
    \label{PNg-pq1-Ne02}
\end{figure*}

\begin{figure}[htb]
    \includegraphics[width=0.48\textwidth]{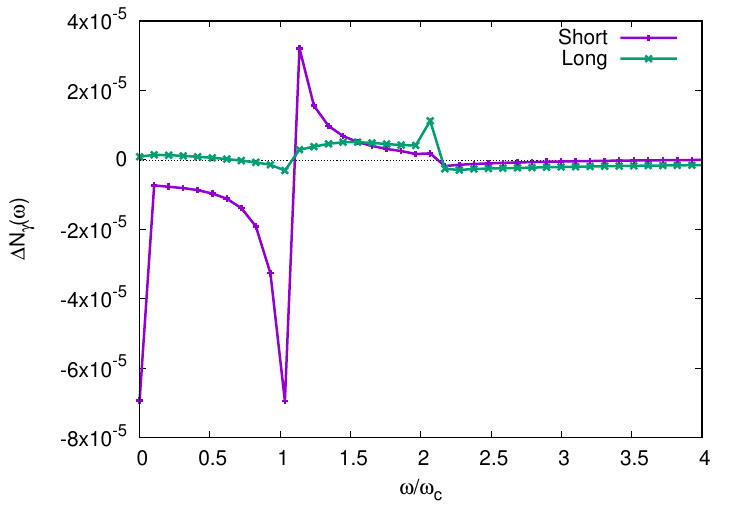}
    \caption{The difference of the Fourier power spectra for the mean photon numbers,
             $\Delta N_\gamma (\omega ) = N_\gamma (\omega; V_t/\hbar\omega_c = +0.1 ) -
             N_\gamma (\omega; V_t/\hbar\omega_c = -0.1)$,
             for a short, and a long excitation. $pq=1$, $N_\mathrm{e}=2$, $g_\gamma=0.04$, $E_\gamma=0.70$ meV.}
    \label{DPNg-pq1-Ne02}
\end{figure}

Inherent in our periodic system is another symmetry breaking worth analyzing better.
In addition to the particular periodic potential (\ref{Vper}) the external magnetic field
breaks the chiral symmetry. In order to explore this effect we look at two cases where
the electron-photon interaction is not very strong. We start with $N_\mathrm{e}=2$ at
$pq = 1$, $g_\gamma=0.04$, and $E_\gamma=0.70$ meV. We choose to excite the system with a
short or long excitation pulse with $V_t/\hbar\omega_c = \pm 0.1$, so especially for the
short pulse (see Fig.\ \ref{pulses}) we are definitely exciting with opposite chirality.
The corresponding Fourier power spectra are displayed in Fig.\ \ref{PNg-pq1-Ne02}.
In the left panel we have the Fourier power spectrum for the mean photon number $N_\gamma(\omega)$
for the short pulse, and in the right panel for the long pulse, in both cases for the two
opposite chiral excitation.
We stress that for clarity we use different vertical scales in the two panels of Fig.\ \ref{PNg-pq1-Ne02}.
We see that the paramagnetic peaks for both pulse lengths are of similar height,
but the diamagnetic peaks are lower for the long one.
The effects of the change in chirality for a pulse with definite length look very small
until the difference
$\Delta N_\gamma (\omega ) = N_\gamma (\omega; V_t/\hbar\omega_c = +0.1 ) -
N_\gamma (\omega; V_t/\hbar\omega_c = -0.1)$ is plotted in Fig.\ \ref{DPNg-pq1-Ne02}
for both pulse lengths.
Fig.\ \ref{PNg-pq1-Ne02} shows a clear, but small, difference in the paramagnetic peak
due to the chirality of the short excitation pulse, and not much can be said about the diamagnetic
peak. For the long excitation pulse the difference $\Delta N_\gamma (\omega )$ is much smaller as could
be expected as the chirality difference is small for the long pulse for $V_t/\hbar\omega_c = \pm 0.1$.

\begin{figure*}[htb]
    \includegraphics[width=0.48\textwidth]{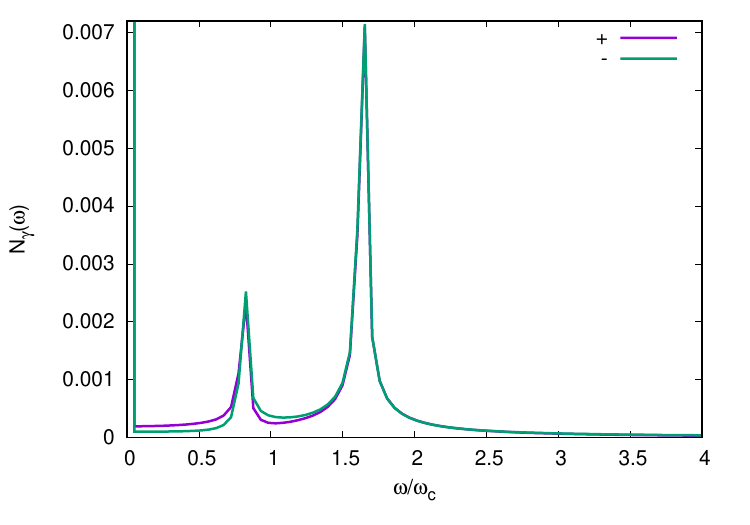}
    \includegraphics[width=0.48\textwidth]{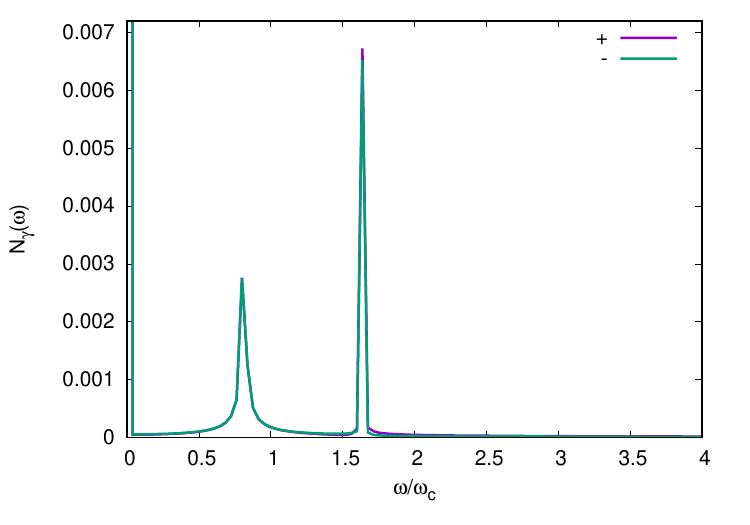}
    \caption{The Fourier power spectra for the mean photon number $N_\gamma (\omega )$
             for $V_t/\hbar\omega_c = \pm 0.1$
             for a short (left), and a long excitation (right). $pq=2$, $N_\mathrm{e}=2$, $g_\gamma=0.04$,
             $E_\gamma=1.00$ meV.}
    \label{PNg-pq2-Ne02}
\end{figure*}

\begin{figure}[htb]
    \includegraphics[width=0.48\textwidth]{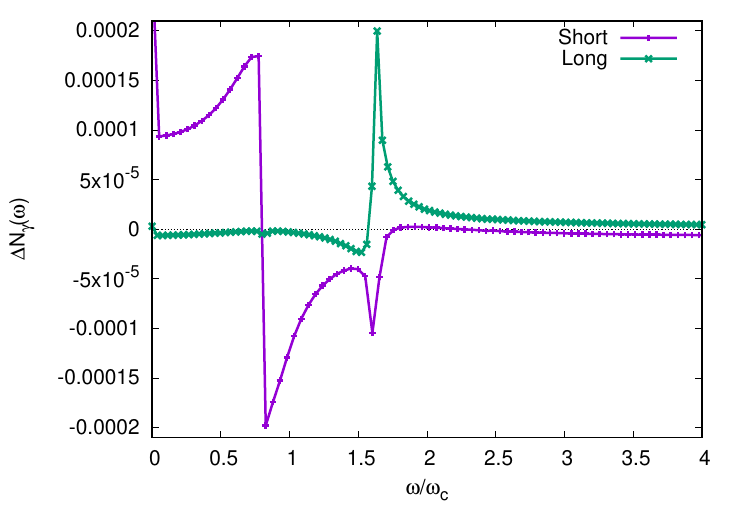}
    \caption{The difference of the Fourier power spectra for the mean photon number
             $\Delta N_\gamma (\omega ) = N_\gamma (\omega; V_t/\hbar\omega_c = +0.1 ) -
             N_\gamma (\omega; V_t/\hbar\omega_c = -0.1)$
             for a short, and a long excitation. $pq=2$, $N_\mathrm{e}=2$, $g_\gamma=0.04$,
             $E_\gamma=1.00$ meV.}
    \label{DPNg-pq2-Ne02}
\end{figure}

The results for the difference in the excited modes for excitations with opposite chirality
at $pq=1$ open the question of how does this look at the higher magnetic field corresponding
to $pq=2$, but with all other parameters unchanged, except for the photon energy now
chosen as $E_\gamma = 1.00$ meV. Fig.\ \ref{PNg-pq2-Ne02} shows the Fourier power spectra for
$N_\gamma(\omega)$ for both chiralities and the short (left) and long pulse (right).
Now the vertical scales of the subfigures are the same and only a small difference between the
results for different chirality are seen, so again we present the difference
$\Delta N_\gamma (\omega ) = N_\gamma (\omega; V_t/\hbar\omega_c = +0.1 ) -
N_\gamma (\omega; V_t/\hbar\omega_c = -0.1)$ in Fig.\ \ref{DPNg-pq2-Ne02}.

\begin{figure*}[htb]
    \includegraphics[width=0.48\textwidth]{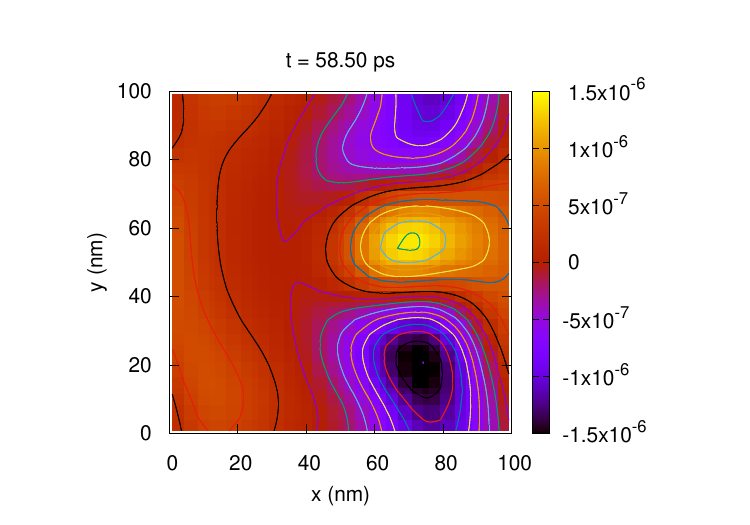}
    \includegraphics[width=0.48\textwidth]{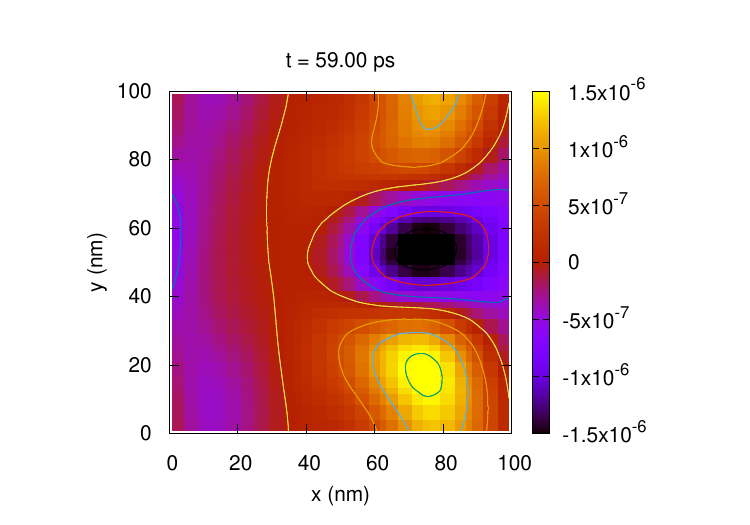}\\
    \includegraphics[width=0.48\textwidth]{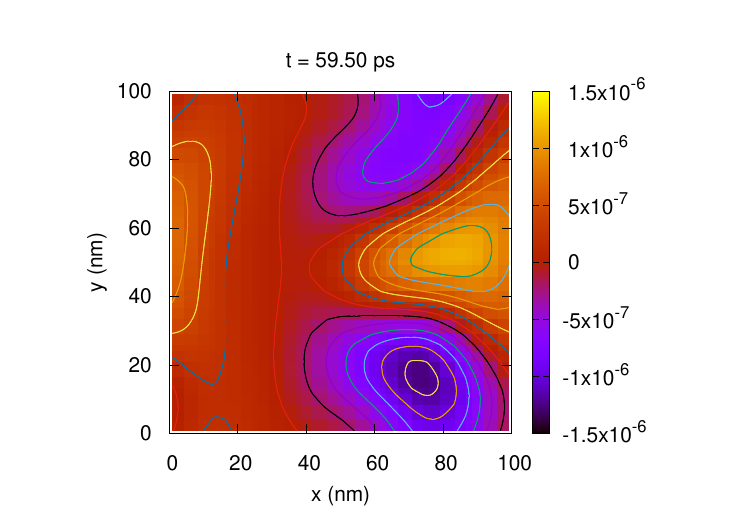}
    \includegraphics[width=0.48\textwidth]{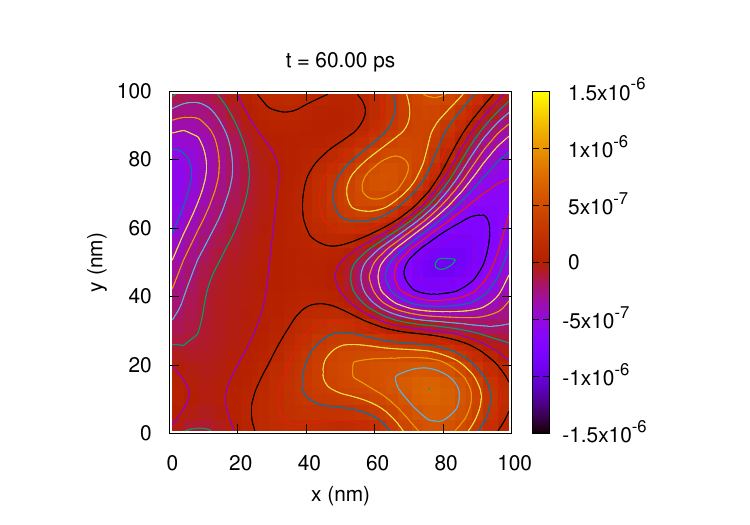}
    \caption{The change in electron density $\Delta n_\mathrm{e} = n_\mathrm{e}(t)-n_\mathrm{e}(0)$
             for 4 different times well after the excitation pulse has vanished.
             $pq=2$, $N_\mathrm{e}=3$, $E_\gamma =1.0$ meV, $g_\gamma =0.08$.
             Long excitation pulse with $\hbar\omega_\mathrm{ext} = 3.5$ meV, and $\hbar\Gamma = 0.5$ meV. }
\label{Ind-ne}
\end{figure*}

As before the main difference for the short excitation pulse is found for the
paramagnetic peak, but in addition there is a smaller change for the diamagnetic
peak, that this time around can not be ignored. For the long excitation pulse only a
small difference is found for the diamagnetic peak. It is clear why the paramagnetic peak
changes with chirality as the electron-photon paramagnetic interaction (\ref{e-g})
is an integral of the inner product of the electron current density and the vector field
of the photons $\bm{A}_\gamma$ and, for example, for the short excitation pulse the
modulation of the interaction either adds or subtracts to the current density.
This modulation is weak ($V_t/\hbar\omega_c = \pm 0.1$) as is reflected in the slight
change induced in the paramagnetic peak. How is the diamagnetic peak influenced by the
chirality of the excitation for the higher magnetic field corresponding to $pq=2$?
To answer this question we have to look at the diamagnetic part of the electron-photon
interaction (\ref{e-g}) and specially how it looks for the particular TE$_{011}$ cylindrical
cavity mode (\ref{Ag})-(\ref{lIxx}). The 2DEG is in a homogeneous external magnetic field
and the electrons are under the influence of the Lorentz force that is larger for $pq=2$
than for $pq=1$. As the electron-photon interaction is modulated the change in the circular
current density leads to a Lorentz force on it that leads to the density being condensed
or expanded depending on the sense of the change in the rotation of the current density.
The diamagnetic electron-photon interaction through the TE$_{011}$ cavity photon mode
depends not only on the number of electrons in a unit cell, but quite sensitively on the
shape of the density (\ref{lIxx}). For a circular density in a symmetric potential in a unit cell
this leads to simple monopole, or breathing, collective oscillations of the electron density.
This effect can be seen in the present array with a unit cell with broken symmetry, albeit
other weaker modes can be observed through the time-dependent induced electron density.
They arise due to the broken symmetry of the unit cell and by the influence of the
square superlattice. In Fig.\ \ref{Ind-ne} the induced electron density is seen for
4 different times close to $t=60.0$ ps for the system with $pq=2$, $N_\mathrm{e}=3$,
$E_\gamma =1.0$ meV, $g_\gamma =0.08$.

These are not neighboring points in time as the time-step for the integration of the
Liouville-von Neumann equation (\ref{L-vN}) is $\Delta t=0.02$ ps, but from the figure we can
identify a clear breathing mode along the minima of $V_\mathrm{per}$. In addition slight
dipole and quadrupole modes are visible together with effects of the external magnetic field.

\section{Summary and Conclusions}
\label{Conclusions}
We use a QED-DFT-TP formalism to explore the effects of broken symmetries on the collective
oscillations of a 2DEG in a homogeneous external magnetic field and an array of unit cells
of a particular type.
The electron system is placed in a cylindrical FIR cavity with a TE$_{011}$ mode.
The external magnetic field breaks the time or the parity symmetry of the system, and the
lack of reflection symmetry of the periodic potential along one of the spatial directions
in the unit cells further complicates the dynamics of the electrons together with the fact
that the vector potential of the cavity photons is spatially dependent.

In the analysis it is essential that the QED-DFT-TP formalism allows for the inclusion of
higher order photon transitions of both the para- and the diamagnetic parts of the
electron-photon interactions and the external magnetic field and the broken spatial symmetry of the
unit cell of the periodic system require a full description of its geometry and shape with
high accuracy in configuration and reciprocal space.

In systems with circular symmetry of the potential in each unit cell, like for an array
of quantum dots or rings, the excitation used here only excites circular currents thus
influencing their dynamical orbital magnetization, their mean photon number and cyclotron
resonances. The broken symmetry of the unit cell in the present model adds center of mass,
and quadrupole density oscillations, but as no linear impulse is imposed on the system
all the collective oscillations appear at energies dictated by the underlying cyclotron
resonances and photon transitions of the TE$_{011}$ cavity mode.
The electron-photon interactions are seen to strongly influence or modify the
center of mass oscillations. For the higher magnetic field, where the influence of the
square lattice is less than for low magnetic field, the photon interaction makes the
center of mass oscillations, or the dipole oscillations, more regular.

We find the influence of chirality of the initial excitation in the
Fourier power spectra of the mean photon number in the system. This effect is small
as we have limited the analysis to weak electron-photon coupling of two electrons
in each unit cell in order to increase their calculational accuracy. We can find this
chiral difference both for the peaks in the Fourier power spectra assigned to
one-photon paramagnetic transitions and two-photon diamagnetic ones. The chirality
difference in the latter ones is caused by the influence of the Lorentz force on the electron
density as the initial excitation modulates the charge currents in the system.

For weak electron-photon interaction paramagnetic Fourier power spectra peaks
for the mean photon number are prominent, but as the electron-photon coupling is increased
the diamagnetic ones dominate. This again underlies the importance of properly accounting
for the diamagnetic electron-photon interaction for electron systems in an external
magnetic field.

The vector potential describing the cavity photon field is spatially dependent.
It makes the electron-photon interaction to promote both electrical and magnetic
transitions. As the excitation pulse directly modulates the amplitude of the
persistent current density of the 2DEG the Lorentz force in turn creates
monopole (and higher order) density oscillations that for the case of a strong external
magnetic field strongly modulates the diamagnetic electron-photon interaction.
This connection between the para- and the diamagnetic electron-photon interactions
is essential in the present electron-cavity system.

Monopole density breathing modes of the system play an important role even when the
symmetry of the potential in unit cells is broken in one spatial direction.
The asymmetry of the periodic superlattice potential hosting the 2DEG prevents
the exchange Coulomb interaction in enhancing the spin splitting of the Landau subbands
when 3 electrons are in a unit cell with two magnetic unit flux quanta flowing through it.
For arrays of quantum dots and rings this was not the case, but instead increased electron-photon
interaction strength would usually reduce this effect
there \cite{PhysRevB.109.235306,PhysRevB.110.205301,PhysRevB.111.115304}.

Though we have concentrated here on the role of the para- and the diamagnetic electron-photon
interaction in the time-dependent system after the excitation pulse has been applied to it,
it is clear that the arguments used here to explain their connection and the importance of
the diamagnetic one apply equally to the static electron-cavity system.

\begin{acknowledgments}
This work was financially supported by the Research
Fund of the University of Iceland Grant No.\ 92199, and
the Icelandic Infrastructure Fund for ``Icelandic Research
e-Infrastructure (IREI)''. The computations were performed
on resources provided by the Icelandic High Performance
Computing Center at the University of Iceland. V.\
Mughnetsyan and V.G.\ acknowledge support by the Higher
Education and Science Committee of Armenia (Grant No.\
24LCG-1C004).
V.\ Moldoveanu acknowledges financial support from the Core Program of the National
Institute of Materials Physics, granted by the Romanian Ministry of Research, Innovation and
Digitalization under the Project No.\ PC2-PN23080202.
W.-H.K.\ acknowledges the support from the National Science and Technology Council, Taiwan, under
Grants No.\ NSTC 113-2221-E-845-00 and No.\ NSTC 114-2221-E-845-002.
H.-S.\ Goan acknowledges support from the National Science and Technology Council (NSTC), Taiwan, under Grants No.\ NSTC 113-2112-M-002-022-MY3, No.\ NSTC 113-2119-M-002-021,  No.\ NSTC 114-2119-M-002-018, No.\ NSTC 114-2119-M-002-017-MY3, from the US Air Force Office of Scientific Research under Award Number FA2386-23-1-4052 and from the National Taiwan University (NTU) under Grants No.\ NTU-CC-114L8950, No.\ NTU-CC-114L895004 and No.\ NTU-CC-114L8517. H.-S.\ Goan is also grateful for the support of the {\lq\lq}Center for Advanced Computing and Imaging in Biomedicine (NTU-114L900702){\rq\rq} through the Featured Areas Research Center Program within the framework of the Higher Education Sprout Project by the Ministry of Education (MOE), Taiwan, the support of Taiwan Semiconductor Research Institute (TSRI) through the Joint Developed Project (JDP) and the support from the Physics Division, National Center for Theoretical Sciences (NCTS), Taiwan.
J.-D.C.\ acknowledges the support from the National Science and Technology Council, Taiwan, under
Grants No.\ NSTC 114-2112-M-002-033 and No.\ NSTC 113-2112-M-002-032. J.-D.C.\ is also grateful for
the support from the Physics Division, National Center for Theoretical Sciences, Taiwan.
\end{acknowledgments}

%----------------------------------------------------------------------------------------
%

\appendix
\section{Technical details in the methodology}
\label{Tech-details}

We use the 8$pq$ lowest Landau subbands, the 10 lowest eigenstates of the photon number operator,
and a $16\times 16$ nonequispaced grid in the first Brillouin zone of the reciprocal space, $\bm{\theta}$,
built on a repeated 4-point Gaussian quadrature. $pq$ is the number of magnetic flux units through
the unit cell of the periodic superlattice, here either 1 or 2. A $13\times 13$ reciprocal lattice is
used, i.e.\ $G_i\in \{-6,\cdots,0,\cdots,6\}$ for $i=1\; \mbox{and}\; 2$.

The ground state, or static, and the dynamical calculations are done in a linear
functional basis constructed as a
tensor product (TP) of electron and photon states
$|\bm{\alpha\theta}\sigma n\rangle = |\bm{\alpha\theta}\sigma\rangle\otimes|n\rangle$.
The electron states were proposed by Ferrari \cite{Ferrari90:4598}, and used by
Gudmundsson \cite{Gudmundsson95:16744} and Silberbauer \cite{Silberbauer92:7355}.
The photon states are the eigenstates of the photon number operator
$N_\gamma = a^\dagger_\gamma a_\gamma$ with eigenvalue $n$. $\sigma$ is the spin label
$\{\uparrow\downarrow\}$, and $\bm{\alpha}$ is a composite quantum number for the Landau subband number.

The self-consistent diagonalization of the static total Hamiltonian leads to the states
$|\bm{\alpha\theta}\sigma n)$, but the Liouville-von Neumann Eq.\ (\ref{L-vN}) is solved in
the $\{|\bm{\alpha\theta}\sigma n\rangle\}$-basis as most matrix elements are known analytically there
\cite{Gudmundsson95:16744,PhysRevB.110.205301}.
%

%----------------------------------------------------------------------------------------
%

%\bibliography{mod_qd}

%apsrev4-2.bst 2019-01-14 (MD) hand-edited version of apsrev4-1.bst
%Control: key (0)
%Control: author (8) initials jnrlst
%Control: editor formatted (1) identically to author
%Control: production of article title (0) allowed
%Control: page (0) single
%Control: year (1) truncated
%Control: production of eprint (0) enabled
%

%--------------------------------------------------------------------------------------
\end{document}